\RequirePackage{lineno}
\documentclass[aps,prb,linenumbers,preprint,groupedaddress,superscriptaddress]{revtex4}
\usepackage{graphicx}
\usepackage{epstopdf}
\usepackage{textcomp}
\begin{document}
\title{Acoustic Kerr nonlinearity of wave propagation in a planar nanoelectromechanical waveguide}

\author{M. Kurosu}
\email{kurosu_megumi_s5@lab.ntt.co.jp; daiki.hatanaka.hz@hco.ntt.co.jp}
\affiliation{NTT Basic Research Laboratories, NTT Corporation, Atsugi-shi, Kanagawa 243-0198, Japan}
\affiliation{Department of Physics, Tohoku University, Sendai 980-8578, Japan}

\author{D. Hatanaka}
\email{kurosu_megumi_s5@lab.ntt.co.jp; daiki.hatanaka.hz@hco.ntt.co.jp}
\affiliation{NTT Basic Research Laboratories, NTT Corporation, Atsugi-shi, Kanagawa 243-0198, Japan}
\author{H. Yamaguchi}

\affiliation{NTT Basic Research Laboratories, NTT Corporation, Atsugi-shi, Kanagawa 243-0198, Japan}
\affiliation{Department of Physics, Tohoku University, Sendai 980-8578, Japan}
%\pagewiselinenumbers

\begin{abstract}
%\begin{linenumbers}
Nonlinearity is the key to introducing novel concepts in various technologies utilizing traveling waves. 
In contrast to the field of optics, where highly functional devices have been developed using optical Kerr nonlinearity\cite{Marin-Palomo2017, Dudley_Supercontinuum_2006, Hansryd_1016354}, such a nonlinear effect in acoustic devices has yet to be fully exploited. 
Here, we show that most fundamental nonlinear phenomena of self-phase modulation (SPM), cross-phase modulation (XPM) and four-wave mixing (FWM) caused by the acoustic Kerr effect are quantitatively characterized using a newly developed platform consisting of a planar nanoelectromechanical waveguide (NEMW). 
Combining the cutting-edge technology of a high crystalline quality NEMW with a piezoelectric interdigital transducer (IDT), we efficiently excite an intense and long-lived traveling wave sufficiently to induce and characterize acoustic nonlinearity.
The observed nonlinear phenomena are precisely described by the model using the nonlinear Schr\"{o}dinger (NLS) equation, so that this architecture enables the nonlinear dynamics to be perfectly tailored. The flexible and integratable platform extends the ability to manipulate acoustic wave propagation on a chip, thus 
offering the potential to develop highly functional devices and study novel nonlinear acoustics.
%\end{linenumbers}
\end{abstract}

\maketitle
Acoustic traveling waves in solid-state structures are capable of coherent energy transfer both in classical and quantum domains on a chip\cite{PRL.121.040501,Hermelin2011, Gustafsson_2014_Science}.
There have been a number of studies with the aim of utilizing acoustic waves for signal processing application such as microwave-to-optical converters and quantum computing\cite{PhysRevApplied.7.024008, Fang2016, kuzyk2018scaling}.
This is because, compared with electromagnetic waves, acoustic devices have the distinct advantages of a short wavelength and small energy loss in on-chip applications.
The key to improving the ability of the on-chip acoustic manipulation is a nonlinear effect, which allows various types of advanced control to be realized including short pulse generation, frequency conversion and amplification, as already demonstrated in nonlinear optics\cite{Agrawal_5th,Hansryd_1016354, PhysRevLett.45.1095, Dudley_Supercontinuum_2006}.

However, compared with work in the optics field, there have been few studies of nonlinear traveling acoustic wave due to the lack of a suitable platform to realize low-loss wave-guiding and a transducer capable of exciting an intense nonlinear wave.
Although some pioneering studies show that nonlinear wave propagation has been observed in solid crystals\cite{PhysRevLett.89.285504, PhysRevB.70.214307, PhysRevB.95.064306}, a strong laser pulse is needed to induce nonlinearity and the experiments must be conducted at low temperature. Such conditions are unsuitable for practical use and monolithic on-chip integration.
In contrast, nonlinear stress-strain relation induced by geometric nonlinearity, has been intensively investigated in nanomechanical resonators.
A number of intriguing results have been reported including mode coupling, a phononic frequency comb, frequency stabilization, and chaos\cite{Foster2016, PhysRevLett.118.033903, Antonio2012, PRB.79.165309_ML}.
Inspired by the development of nanomechanical technology, a novel acoustic platform for a NEMW has recently been realized that is constructed from a suspended semiconductor membrane array. This hosts its excellent properties, such as engineerable dispersion, low propagation loss, design flexibility and semiconductor-based integration. Thus, this architecture has enabled the demonstration of electrical phonon manipulation, energy focusing by dispersion, and the active manipulation of phononic band structures on a chip\cite{Hatanaka_Phonon_2014, Hatanaka_Phonon_2015, Cha2018, NC_Kurosu2018}.

By combining a newly designed 33-mm long acoustic waveguide structure with an IDT that enables the efficient excitation of strong acoustic vibration with a moderate input amplitude, we demonstrate the fundamental nonlinear phenomena of SPM, XPM and FWM induced by the acoustic Kerr effect and confirm that they can be controlled by adjusting the input excitation amplitude and propagation distance.
While optical Kerr effects originate in refractive index variation due to the electric field of light, acoustic Kerr effects are caused by geometric nonlinearity\cite{Lifshitz}.
These nonlinear propagation dynamics can be observed because of the low-loss, single-mode and long transmission channel with a high quality GaAs/AlGaAs single crystal heterostructure.
This platform provides proof of the capacity to manipulate nonlinear acoustic wave propagation on a chip and will pave the way to the development of nonlinear acoustic devices.

A vibrating membrane is fabricated by sacrificially etching an Al$_{0.7}$Ga$_{0.3}$As layer as shown in Fig. 1c, and this is 33-mm long when folded to realize a small device footprint as shown in Fig. 1a. 
This waveguide hosts a continuous transmission band between 2.4 and 7.4 MHz except in the 7.4-7.8 MHz bandgap regime, which is caused by Bragg reflection of a vibration from a periodically arrayed air holes along the waveguide as shown in Fig. 1b.
This phononic crystal structure modulates the group velocity dispersion (GVD) of the device as shown in Fig. 1e, indicating that the dispersive effect can be tuned by engineering the periodic structure or selecting the operating frequency. 
Considering our goal, which is to confirm the ability to control nonlinear wave propagation, the dispersion effect on the propagation dynamics should be minimized. In our work, we chose an IDT electrode pitch of 20 $\mu$m to correspond with the operating frequency of a low GVD regime. Therefore, the IDT can excite large flexural vibrations around 5.38 MHz at which the GVD coefficient $k_2$ is $\sim -2\times10^{-10}$ s$^{2}$ m$^{-1}$ estimated with a finite element method (FEM) simulation as shown in Fig. 1d and 1e. The resultant waves can be measured at various distances in the waveguide by adjusting the laser spot position of an optical interferometer. All the experiments described here were performed at room temperature and in a moderate vacuum ($\sim$ 10 Pa).

The mechanical motion of the suspended vibrating plate is governed by the Euler-Bernoulli equation\cite{Lifshitz}. From this, we derive a wave equation, namely the NLS equation, which can be used to predict the nonlinear dynamics of acoustic wave propagation in a waveguide\cite{Nayfeh_book}. Here, the envelope of the vibrating pulse centered around wavenumber $k$ and the angular frequency $\omega$ was assumed to vary slowly in the temporal and spatial domains. Considering the amplitude $A(x,t)$ where the $x$ is the propagation distance and $t$ is time, the NLS equation is given by,
\begin{eqnarray}
%\frac{\partial A}{\partial x}=-\frac{\alpha}{2}A-\frac{i k_2}{2}\frac{\partial^2 A}{\partial T^2}+\frac{k_3}{6}\frac{\partial^3 A}{\partial T^3}+i \xi |A|^2 A, 
\frac{\partial A}{\partial x}=-\frac{\alpha}{2}A-\frac{i k_2}{2}\frac{\partial^2 A}{\partial T^2}+i \xi |A|^2 A, 
\label{eq:aa}
\end{eqnarray}
where $\alpha$, $k_2=\frac{\partial^2 k}{\partial \omega^2}$ and $\xi$ denote the linear loss, GVD coefficient and an effective nonlinear parameter, respectively.
$T = t-x/v_\mathrm{g}$ is the time in a moving frame where $v_\mathrm{g}$ is group velocity.
In this device, $\alpha$ is set at 0.29 dB mm$^{-1}$, which is determined by a time-of-flight measurement described in detail elsewhere\cite{NC_Kurosu2018}.
In this new structure, the value was greatly improved compared with that in our previous report\cite{Hatanaka_Phonon_2014} because of the process and device design optimization.
It is worth noting that the third term on the right hand side of equation (1) contains $|A|^2$ representing the squared amplitude. This governs the acoustic Keff nonlinearity and cannot be negligible when the intensity of the traveling wave is significantly large and induces third-order nonlinearity in the system. As a result, the phase of the wave is modulated during propagation, which is known as SPM, especially in the field of nonlinear optics\cite{stolen1978SPM}. When the GVD effect can be ignored, the maximum phase shift $\theta_\mathrm{max}$ caused by SPM is written as\cite{Agrawal_5th},
\begin{eqnarray}
\theta_\mathrm{max}\left(x \right)  = \xi |A\left(x, 0 \right) |^2 x_\mathrm{eff}
\end{eqnarray}
where $x_\mathrm{eff}$ is the effective distance, which is defined by $x_\mathrm{eff} = \frac{1-\exp(-\alpha x)}{\alpha}$. As seen from equation (2), the SPM-induced phase shift is proportional to the instantaneous squared amplitude of the pulse, and thus its temporal response is followed by the pulse envelope as shown in Fig. 2a. The sign of the nonlinear parameter $\xi$ determines the polarity of the SPM. Additionally, the phase is accumulated while propagating in the waveguide. Our high crystalline quality NEMW with a transmission channel long enough to allow us to characterize the phase accumulation is a suitable platform on which to observe this nonlinear dynamics. 

The SPM process is at the heart of nonlinear phenomena and thus, it is of prime importance to investigate the effect. To that end, an intense pulsed acoustic wave is efficiently excited using the IDT electrode located at one end of the waveguide, where a Gaussian-shaped pulse $A(0,T)=A_0 \exp⁡\left(-\frac{T^2}{2T_0^2} \right) $ is used as the input. The time evolution of the amplitude and phase of the pulse is measured at distances of 0, 10, and 19.5 mm from the IDT as shown in Fig. 2b-2d and 2e-2g,  respectively. The amplitude of a pulse envelope of width $T_0= 40\ \mu$s increases as the excitation voltage is increased to 1.2 V$_\mathrm{rms}$ at $x = 0$ mm (Fig. 2b), whereas the temporal response of the phase is nearly flat and is invariant even when the excitation voltage is changed (Fig. 2e). 
However, as the wave propagates, the phase is being modulated through the SPM process, and it finally reaches -50 degrees at $x = 19.5$ mm with a 1.2 V$_\mathrm{rms}$ excitation voltage (Fig. 2g). 
This negative phase shift due to the sign of $\xi$ becomes apparent at larger excitation amplitudes. 
We note that the negative phase shift indicates the leading phase because 
we used the conventions of $\exp(-\mathrm{i}\omega t)$ as a fundamental wave\cite{Nayfeh_book}.
Furthermore, to validate these results, we simulate the propagation distance dependence of the phase shift at various voltages using equation (1) as shown in Fig. 2h (see also Methods). The experimentally observed variation in the phase can be well reproduced by a simulation that involves applying the split-step Fourier method to the NLS equation (see Supplementary Information for details), where the larger SPM occurs at a longer distance and a larger amplitude. From this comparison, the effective nonlinear parameter of this NEMW is obtained as $\xi = -16$ nm$^{-2}$ m$^{-1}$. %Then, all the parameters except $k_2$ in equation (1) are experimentally determined, allowing us to quantitatively predict and control nonlinear wave dynamics traveling in this system.
This value is sufficiently large to observe novel nonlinear phenomena.
For example, acoustic soliton compression is also possible with realistic parameters as shown in Supplementary Information. 
%Since the experimental value of $k_2$ is fluctuated as showin in Fig. 1e, the FEM-simulated value is used in the simulation.

Although SPM works only with a single acoustic wave, acoustic Kerr nonlinearity can also be exploited to tune and manipulate the propagation dynamics through the interaction between different acoustic waves. An incident wave can interact with a strongly-excited wave through this third-order nonlinear effect, thus resulting in the phase of the incident wave being modulated. 
This nonlinear interaction can generate a new acoustic wave, whose phenomenon is known as FWM. 
These effects have been intensively studied in the field of nonlinear optics leading to the development of a range of light manipulation techniques including
wavelength conversion, amplification, supercontinuum generation and solitons\cite{Agrawal_5th,Hansryd_1016354, PhysRevLett.45.1095, Dudley_Supercontinuum_2006}.
Figure 3a-3f show the spectral response of a continuous wave (signal) and an intense pulsed wave (pump) excited from the IDT at $f_\mathrm{s}$ = 5.264 MHz and $f_\mathrm{p}$ = 5.383 MHz at various positions in the waveguide as shown in Fig. 1a. 
As described above, the strong pump wave activates the nonlinearity and this induces a temporal phase shift in the signal wave, which is called XPM. 
Hence, the temporally varying phase modifies the instantaneous frequency of the pulse, resulting in the bottom of the signal spectrum being broadened. 
On the other hand, under appropriate conditions that satisfy energy and momentum conservation, the FWM process also permits the generation of a new wave, called an idler.
The output spectra reveal the idler wave at $f_\mathrm{i}$ = 5.502 MHz, which satisfies the energy conservation requirement $2f_\mathrm{p} = f_\mathrm{s}+f_\mathrm{i}$. 
The momentum conservation requirement $2k_\mathrm{p}=k_\mathrm{s}+k_\mathrm{i}$ is largely met due to the small GVD and the frequency separation between the waves involved, namely $k_\mathrm{p} \sim k_\mathrm{s} \sim k_\mathrm{i}$ where $k_\mathrm{p}$, $k_\mathrm{s}$ and $k_\mathrm{i}$ are the wavenumber of the pump, signal and idler,  respectively. The spectral responses resulting from the XPM and FWM processes can be numerically calculated using equation (1) with the parameters obtained from the SPM experiment, as shown by the solid blue lines in Fig. 3a-3f. The numerically calculated and experimental spectra are good agreement, where the spectral broadening in the signal via XPM and the idler generation via FWM are well reproduced.

To further elucidate the nonlinear acoustic process, the propagation dynamics of the signal, pump, idler and broadened signal due to XPM are investigated by measuring the amplitude of each wave at distances of 0 to 31 mm as shown in Fig. 4. At the beginning of the propagation, the idler peak amplitude (green) increases greatly and it reaches its maximum value of $\sim$ 20 pm at $x$ = 7 $\sim$ 10 mm. However, the pump (red) and signal (blue) are attenuated due to the loss, which suppresses the efficiency of the FWM process. 
Additionally, the idler experiences energy dissipation during propagation. Therefore, idler generation is overwhelmed by the loss, resulting in the output amplitude being reduced. The dynamics of nonlinear wave propagation is quantitatively captured by a numerical calculation based on the NLS equation (solid lines). The results provide proof that the transmission properties of the signal, pump and idler waves originate from the nonlinear effects of the SPM, XPM and FWM processes. To the best of our knowledge, ours is the first report of the experimental demonstration and perfect modeling of SPM, XPM and FWM using a solid acoustic waveguide. 

In conclusion, we demonstrated SPM, XPM and FWM resulting from the acoustic Kerr effect in a planar NEMW and confirmed the effectiveness of using the NLS equation to investigate the acoustic wave dynamics. Combined with the phononic crystal property, which enables the GVD to be engineered, the novel acoustic platform offers us the ability to generate and fully manipulate a range of nonlinear phenomena including solitons and supercontinuum generation. This will enhance the availability of acoustic waves and phonons for use in the fields of acoustic signal processing and opto-/electro-mechanics as well as investigations into fundamental nonlinear physics.
\\\\
{\bf Methods}\\
\noindent{\bf Measurement setup.}
The NEMW is excited by applying an alternating voltage or Gaussian-shaped voltage from a function generator (NF WaveFactory 1974). The resultant flexural vibrations are measured with a laser Doppler interferometer (Melectro V1002). The demodulated electric signals are sent to a vector signal analyzer (HP89410A) in the spectral measurements as shown in Figs. 1d and 3. The signals are also sent to a lock-in amplifier (Zurich Instruments UHFLI) where in-phase ($X$) and out-of-phase ($Y$) components with respect to the reference frequency $\omega$ are detected using a phase-sensitive detection method. Then, both signals are passed through low-pass filter with a bandwidth of 470 kHz and a time constant of 102.6 ns, and then measured with an oscilloscope (Agilent DSO6014A) for the temporal measurements shown in Figs. 1e and 2b-2g. As a result, the amplitude and phase of the pulse can be estimated from $\sqrt{X^2+Y^2}$ and $\tan^{-1}(Y/X)$, respectively.\\\\

\noindent{\bf Data availability.}
The data that support the findings of this study are available from the corresponding author in response to a reasonable request.

\nolinenumbers
\bibliographystyle{naturem}

\bibliography{Reference}

\vspace{5mm}
\noindent{\bf Acknowledgments}\\
We acknowledge the stimulated discussion in the meeting of the
Cooperative Research Project of the Research Institute of Electrical
Communication, Tohoku University.
This work is partly supported by a MEXT Grant-in-Aid for Scientic Research on Innovative Areas ``Science of hybrid quantum systems" (Grant No. JP15H05869 and JP15K21727).\\\\
{\bf Author Contributions}\\
M.K. performed the measurements, the data analysis and the simulation, and fabricated the device with the help of D.H. and H. Y. M.K., D.H. and H.Y. wrote the manuscript and H.Y. planned the project.

\begin{figure*}[!hbtp]
	\begin{center}
		\includegraphics[scale=0.8]{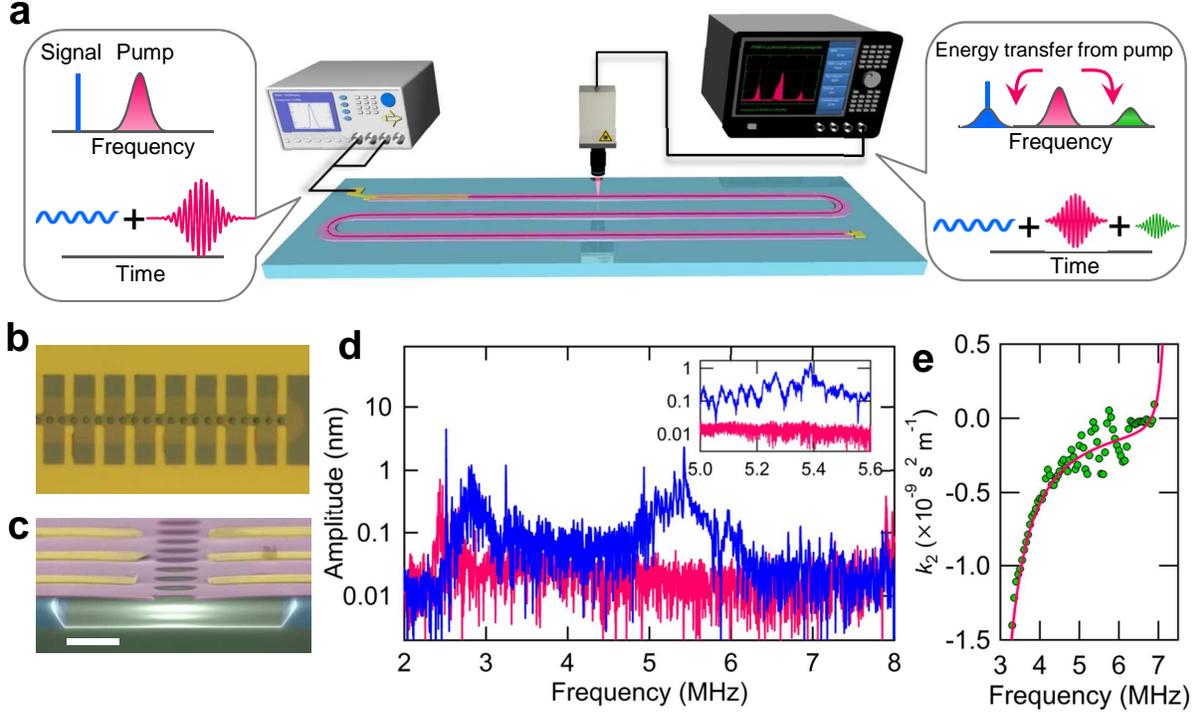}
		\caption{Nanoelectromechanical waveguide. \textbf{a}  A schematic showing the device and the measurement setup. Radio-frequency signals are sent from a function generator to the IDT electrode, where a single pulsed wave, or a continuous wave and pulsed waves (signal and pump) are injected. 
		The left and right insets show the input and output signal configurations in an FWM experiment (see Figs. 3 and 4).
		\textbf{b} Microscope photograph of the IDT electrode and suspended membrane. The width and pitch of the IDT are 5 and 20 $\mu$m, respectively. The width and air-hole pitch of the waveguide are 27.7 and 8 $\mu$m, respectively.
		\textbf{c} A false-colored SEM image of a cross-section with IDT electrodes (yellow). The GaAs membrane with a thickness of 200 nm (pink) is suspended by etching an Al$_{0.7}$Ga$_{0.3}$As sacrificial layer (blue). The scale bar is 5 $\mu$m. 
		\textbf{d} The frequency responses of the device excited by the IDT (blue) and a conventional piezotransduer (red) with 1.0 V$_\mathrm{rms}$, where they are measured at distances $x$ = 5 and 1 mm from each transducer, respectively. The IDT hosts narrow excitation,
		%because the 100-finger electrode array in the IDT restricts the excitation frequency range
		whereas the activated vibration is larger than that from the usual transducer of around 5.4 MHz. The inset shows an enlarged view from 5.0 MHz to 5.6 MHz.
		\textbf{e} GVD coefficient $k_2$ as function of frequency. The experimental results (green dots) are estimated from time-of-flight measurements and smoothed by Gaussian smoothing. The red line is calculated with the FEM simulation, where the internal stress between the GaAs and Al$_{0.7}$Ga$_{0.3}$As layers is included\cite{Liu_2011}.
		} 
		\label{fig 1}
	\end{center}
\end{figure*}

\begin{figure*}[!t]
	\begin{center}
		\includegraphics[scale=0.8]{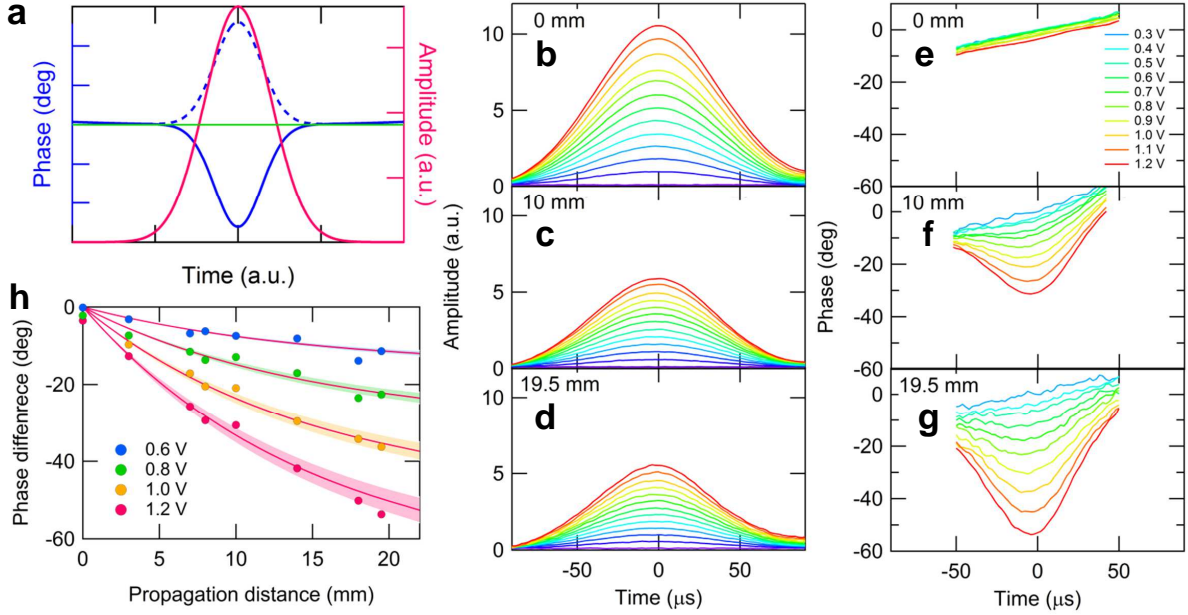}
		\caption{Self-phase modulation. \textbf{a} Phase change of a Gaussian-shaped pulse (red solid line) due to SPM. The positive (negative) nonlinear parameter $\xi$ activates a positive (negative) phase shift as shown by the dashed (solid) blue line. The solid green line denotes the phase of an input pulse. 
		\textbf{b-g} Temporal response of the amplitude (\textbf{b-d}) and phase (\textbf{e-g}) of an acoustic wave pulse measured at a distance $x$ = 0 mm (\textbf{b, e}), 10 mm (\textbf{c, f}), and 19.5 mm (\textbf{d, g}) when excited by the IDT with voltages ranging from 0 V$_\mathrm{rms}$ to 1.2 V$_\mathrm{rms}$. As the amplitude is small around the leading and trailing edges in the pulse, the phase fluctuates and is not determined. Therefore, it is removed from the data. There is a slight slope in the time evolution of the phase on the baseline of the pulse. This might be frequency chirp caused by dispersion effect induced by the periodic electrode array in the IDT.
		\textbf{h} The propagation length dependence of the SPM induced the maximum phase difference with respect to the phase of the pulse with 0.3 V$_\mathrm{rms}$. The solid lines show numerical simulation results for $\xi=-16$ nm$^{-2}$ m$^{-1}$ using the NLS equation. The colored bands indicate the numerical simulation results for $\xi$ = $-15$ $\sim$ $-17$ nm$^{-2}$ m$^{-1}$.
		}
	\end{center}
\end{figure*}

\begin{figure*}[!t]
	\begin{center}
		\includegraphics[scale=0.86]{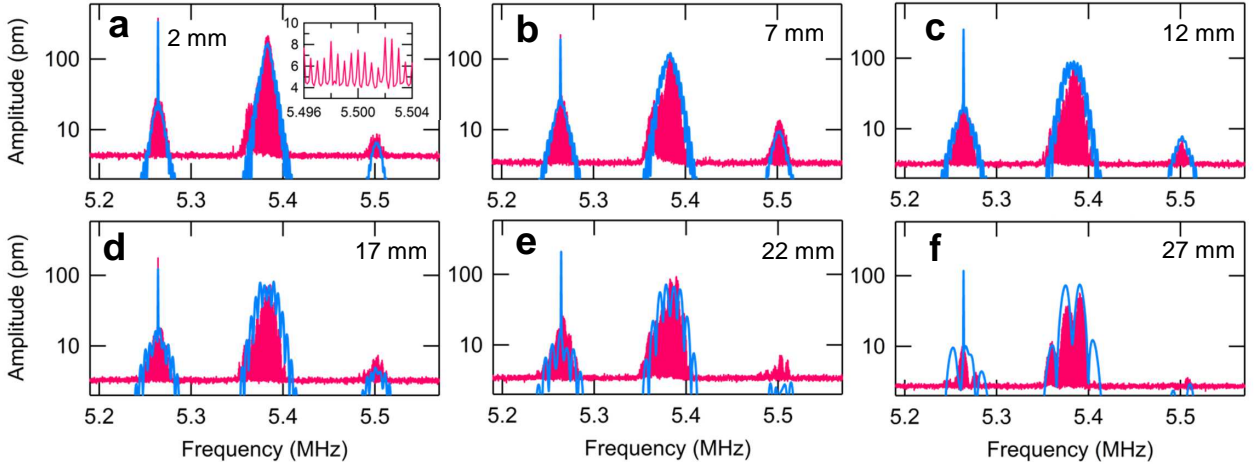}
		\caption{Four-wave mixing. 
			\textbf{a-f} The spectral response of the NEMW measured at distances $x$ = 2 mm (\textbf{a}), 7 mm (\textbf{b}), 12 mm (\textbf{c}), 17 mm (\textbf{d}), 22 mm (\textbf{e}) and 27 mm (\textbf{f}) when exciting a continuous wave signal at 5.264 MHz with an amplitude of 0.4 V$_\mathrm{rms}$ and a Gaussian-shaped pump pulse at 5.383 MHz with a peak amplitude of 2.0 V$_\mathrm{rms}$ and a temporal width $T_0$ = 40 $\mu$s. Spectral broadening occurs in the signal caused by the XPM process. A pulsed idler wave is generated via FWM at around 5.502 MHz. The inset is an enlargement of the figure in (a) showing that equally distant peaks corresponding to a pulse repetition rate of 500 Hz are formed due to the resolution bandwidth of the vector signal analyzer (10 Hz) which is smaller than the pulse repetition rate.
		}
		\label{}
	\end{center}
\end{figure*}

\begin{figure*}[!t]
	\begin{center}
		\includegraphics[scale=0.75]{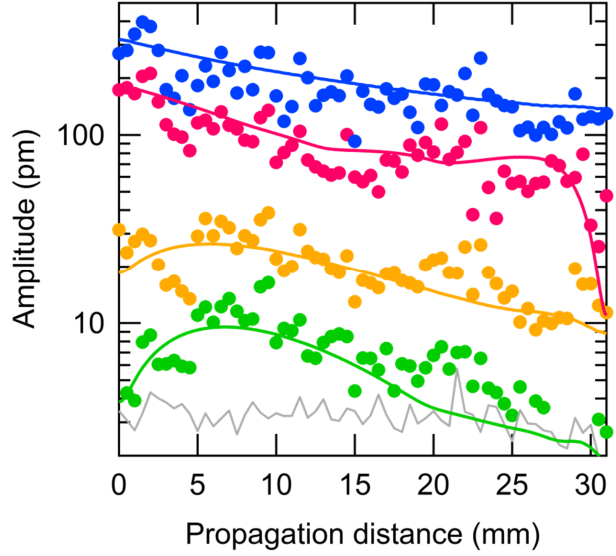}
		\caption{Dynamics of nonlinear wave propagation in the NEMW. The propagation distance dependences of the peak amplitude of the signal, pump, idler and broadened signal due to XPM are plotted as blue, red, green and orange circles respectively. 
		The solid lines indicate the simulation results and the gray line shows the average noise floor of the system. 
		In the simulation, the input spectral amplitude of the signal, pump, idler and broadened signal due to XPM are 320, 180, 3 and 16 pm, respectively.
		}
	\end{center}
\end{figure*}

\end{document}